\documentclass[
aps,
nofootinbib,
prd,
superscriptaddress,
tightenlines,
notitlepage,
twocolumn,
showpacs,
floatfix
]{revtex4-1}

\usepackage{graphicx}
\usepackage{subfigure}
\usepackage{dcolumn}
\usepackage{bm}

\usepackage[colorlinks]{hyperref}
\usepackage{tabularx}
\usepackage{graphicx}
\usepackage{amssymb,bm,tensor}
\usepackage{xcolor}
\usepackage{amsmath}
\usepackage[varg]{txfonts}
\usepackage{enumerate}
\usepackage[normalem]{ulem}
\usepackage{bm}
\usepackage[T1]{fontenc}
\usepackage{orcidlink}

\usepackage{scalerel,stackengine}
\stackMath
\newcommand\reallywidehat[1]{%
	\savestack{\tmpbox}{\stretchto{%
			\scaleto{%
				\scalerel*[\widthof{\ensuremath{#1}}]{\kern-.6pt\bigwedge\kern-.6pt}%
				{\rule[-\textheight/2]{1ex}{\textheight}}
			}{\textheight}%
		}{0.5ex}}%
	\stackon[1pt]{#1}{\tmpbox}%
}

\newcommand{\rmd}{\mathrm{d}}

\begin{document}
	\title{Chaotic motion of particles around a Schwarzschild black hole in a swirling electromagnetic background}
	\author{Wenbin Li \orcidlink{0009-0008-8127-6928}}
	\affiliation{Department of Physics, College of Sciences, Shanghai University, 99 Shangda Road, 200444 Shanghai, China}
	
	\author{Yu-Qi Lei \orcidlink{0000-0002-5629-3071}}
	\email{yuqi_lei@shu.edu.cn}\thanks{Corresponding author}
	\affiliation{Department of Physics, College of Sciences, Shanghai University, 99 Shangda Road, 200444 Shanghai, China}
	\affiliation{Department of Mathematics, College of Sciences, Shanghai University, 99 Shangda Road, 200444 Shanghai, China}
	
	\author{Xian-Hui Ge \orcidlink{0000-0001-6228-0376}}
	\email{gexh@shu.edu.cn}\thanks{Corresponding author}
	\affiliation{Department of Physics, College of Sciences, Shanghai University, 99 Shangda Road, 200444 Shanghai, China}
	
	\begin{abstract}
		We investigate the particle motion around a Schwarzschild black hole immersed in a swirling Bertotti-Robinson-Bonnor-Melvin background. This spacetime provides a physically well-motivated framework for studying how the two different electromagnetic components and the swirling deformation affect particle dynamics near compact objects. By employing Poincar\'{e} sections, the maximum Lyapunov exponent, the Fast Lyapunov indicator, recurrence analysis and bifurcation diagrams, we show that chaotic motion can already appear in the non-swirling Schwarzschild-Bertotti-Robinson black hole. This indicates that the swirling background is not a necessary condition for chaos in this family of spacetimes, it mainly shifts the parameter region where chaos occurs. We further find that the effects of the two electromagnetic fields are very complicated. In particular, the existence of bound orbits is strongly restricted by the strengths of the two electromagnetic fields and their relative direction. These results provide rich numerical evidence that the chaotic motion of particles is associated with the nonlinear interaction between the accessible phase space, the electromagnetic backreaction and the swirling deformation. 
	\end{abstract}
	
	\maketitle
	
	\section{Introduction}\label{sec:1}
	
	Chaos is an unpredictable behavior in nonlinear dynamical systems. Its most notable feature is the sensitive dependence on initial conditions. Two trajectories starting from very close initial positions may separate rapidly during the time evolution, making the long-term motion extremely difficult to predict. In Hamiltonian systems, the emergence of chaos is related to the destruction of integrability, namely, the loss of a sufficient number of independent conserved quantities needed for a complete separation of the equations of motion \cite{Ott:2002cds,Rasband:2015cdns}. Black hole spacetimes play an important role in the study of chaos, because the strong gravitational fields and nonlinear couplings can significantly amplify the deviation from the regular motion. Although Einstein's field equations are highly nonlinear, several standard black hole solutions in General Relativity still possess hidden symmetries to guarantee the integrability of geodesic motion. It is well-known in the Kerr spacetime, where the existence of the Carter constant ensures the separability of the geodesic equations \cite{Carter:1968rr}. However, once the background geometry is deformed by external fields or by other complicated structures, these symmetries may be partially or completely broken, and chaotic motion will appear \cite{Bombelli:1991eg,Letelier:1996he,Santoprete:2001wz,Hashimoto:2016dfz,Dalui:2018qqv,Dalui:2026msw}. 
	
	Chaotic dynamics has been widely investigated in gravitational systems involving black holes and other compact objects. Representative examples include neutral or charged particles moving in perturbed or deformed black hole spacetimes \cite{Karas:1992,Dettmann:1994dj,Sota:1995ms,Vieira:1996zf,Aguirregabiria:1996vq,Gueron:2002jt,Dubeibe:2007hba,Semerak:2010lzj,Contopoulos:2011dz,Semerak:2012dx,Sukova:2013jxa,Lukes-Gerakopoulos:2014dpa,Witzany:2015yqa,Chen:2016tmr,Li:2018wtz,Polcar:2019wfi,Bera:2021lgw,Sun:2021oxg,Zhang:2021uaz,Deich:2022vna,Yang:2023bkq,Cao:2024ihv,Cao:2024pdb,Xu:2025lzd,Lu:2025hky,Cao:2025qpz,Lu:2026kcm}, the motion of spinning particles  \cite{Suzuki:1996gm,Hartl:2002ig,Hartl:2003da,Kao:2004qs,Han:2008zzf,Verhaaren:2009md,Lukes-Gerakopoulos:2016bup,Zelenka:2019nyp,Yuan:2026uyk,Shahzadi:2026dqq}, and geodesic motion in black ring and black string backgrounds \cite{Frolov:1999pj,PandoZayas:2010xpn,Igata:2010cd,Ma:2014aha,Ma:2019ewq,Igata:2020dow,Ma:2022tvs}. Chaos has also been discussed for scalar particles nonminimally coupled to curvature invariants \cite{Wang:2016wcj, Zhou:2020fvd, Zhang:2023lrt}, as well as in black hole thermodynamics in the extended phase space by means of the Melnikov method \cite{Chabab:2018lzf,Mahish:2019tgv,Lei:2020clg,Lei:2024qpu}. These studies indicate that chaos tends to emerge once the integrable structure of the spacetime is sufficiently weakened. They also show that the particle dynamics can serve as a sensitive probe for the background geometry and external fields. 
	
	There is a large number of observational evidence suggests that the supermassive black hole at the center of the Milky Way is surrounded by electromagnetic fields \cite{Eatough:2013nva,EventHorizonTelescope:2019dse,EventHorizonTelescope:2021srq,EventHorizonTelescope:2022wkp}. Recently, an exact black hole solution in Einstein-Maxwell theory was constructed by applying the Harrison transformation and the Ehlers transformation to the Schwarzschild-Bertotti-Robinson spacetime \cite{Podolsky:2025tle}. The resulting solution describes a Schwarzschild black hole immersed in a swirling electromagnetic background formed by the superposition of the Bertotti-Robinson and the Bonnor-Melvin universe \cite{Astorino:2025lih}. The two homogeneous electromagnetic fields are incorporated into a single framework, which makes it possible to study their nonlinear interaction on the black hole spacetime. Therefore, it is interesting to investigate the geodesic motion of particles around this spacetime. In contrast to the Ernst-Schwarzschild geometry, which only involves a single Melvin-type background \cite{Karas:1992,Li:2018wtz}, the present spacetime contains two independent electromagnetic structures. Moreover, unlike the dyonic Kerr-Newman black hole in a Melvin-swirling universe \cite{Cao:2025qpz}, the central object considered here has no intrinsic spin or charge, the stationary and axisymmetric character of the new spacetime is generated entirely by the swirling universe. In this sense, the spacetime provides a useful framework for examining more directly how different electromagnetic backgrounds and the additional swirling parameter affect the integrability of particle motion. 
	
	The purpose of this work is to explore the chaotic motion of particles around a Schwarzschild black hole immersed in a swirling Bertotti-Robinson-Bonnor-Melvin background. Our main goal is not only to identify chaotic trajectories, but also to clarify the respective roles of the external fields. In particular, we show that chaotic motion can already appear in the non-swirling limit, including the more special Schwarzschild-Bertotti-Robinson spacetime. This result implies that swirling is not a necessary condition for chaos in this family of backgrounds. Its main role is to enrich the phase space structure and change the parameter regions in which chaotic, regular and escaping orbits occur. 
	
	This paper is organized as follows. In Sec.~\ref{sec:2}, we review the spacetime metric and discuss its basic physical properties, together with the equations of motion for neutral particles. In Sec.~\ref{sec:3}, we analyze the chaotic dynamics by means of the Poincar\'{e} sections, maximum Lyapunov exponent, Fast Lyapunov indicator, recurrence plots and bifurcation diagrams, considering both the non-swirling and swirling cases. Finally, we present our conclusions and discussions in Sec.~\ref{sec:4}.

	\section{The black Hole Spacetime and Equations of Motion}\label{sec:2}

    In this section, we first briefly review the spacetime geometry of the Schwarzschild black hole embedded into a swirling Bertotti-Robinson-Bonnor-Melvin background. We use natural units $G = c = \hbar = 1$, the action of the Einstein-Maxwell theory is 
    \begin{align}\label{eq:act}
        \mathcal{I} \left[ g_{\mu \nu}, A_\mu \right] =  - \frac{1}{16 \pi} \int_{\mathcal{M}} \rmd^4 x \sqrt{-g} \left( R - F_{\mu \nu} F^{\mu \nu} \right),
    \end{align}
	where $R$ is the Ricci scalar and the Faraday electromagnetic tensor is defined as $F_{\mu \nu} = \nabla_\mu A_\nu - \nabla_\nu A_\mu$. Varying Eq.~\eqref{eq:act} with respect to the metric tensor $g_{\mu \nu}$ and the electromagnetic potential $A_\mu$, respectively, gives the Einstein-Maxwell field equations \cite{Astorino:2025lih}
    \begin{align}
        &R_{\mu \nu} - \frac{1}{2} g_{\mu \nu} R  = 2 \left( F_{\mu \rho} {F_\nu}^\rho - \frac{1}{4} g_{\mu \nu} F_{\rho \sigma} F^{\rho \sigma} \right), \label{eq:EM1}\\
        &\partial_\mu \left( \sqrt{-g} F_{\mu \nu} \right) = 0. \label{eq:EM2}
    \end{align}
    
    For a stationary and axisymmetric electrovacuum spacetime, it is convenient to use the Weyl-Lewis-Papapetrou form \cite{Stephani:2003ese, Griffiths:2009ese}
    \begin{align}\label{eq:WLP}
        \rmd s^2 = - f \left( \Delta_\varphi \rmd \varphi - \omega \rmd t \right)^2 + f^{-1} \left[ \rho^2 \rmd t^2 - e^{2 \gamma} \left( \rmd \rho^2 + \rmd z^2 \right) \right],
    \end{align}
    together with an electromagnetic one-form potential
    \begin{align}\label{eq:em_pot}
        A_\mu \rmd x^\mu = A_t \rmd t + A_\varphi \rmd \varphi.
    \end{align}
    
    To solve the Einstein-Maxwell field equations, we introduce the complex electromagnetic Ernst potential $\bm \Phi$ and the gravitational Ernst potential $\bm {\mathcal{E}}$ \cite{Ernst:1967by}
    \begin{align}\label{eq:Ernst_pot}
        \bm{\Phi} \equiv A_\varphi + i \tilde A_t, \qquad \bm{\mathcal{E}} \equiv f - \bm\Phi \bm\Phi^* - i h,
    \end{align}
    where $\tilde A_t$ and $h$ are defined through
    \begin{align}\label{eq:gra}
        &\vec{\nabla} \tilde A_t = f \rho^{-1} \vec{e}_\varphi \times \left( \vec{\nabla} A_t + \omega \vec{\nabla} A_\varphi \right), \\
        &\vec{\nabla} h = -f^2 \rho^{-1} \vec{e}_\varphi \times \vec{\nabla} \omega - 2 \operatorname{Im} \left( \bm\Phi^* \vec{\nabla} \bm\Phi \right). 
    \end{align}
    Here $\vec{\nabla}$ denotes the gradient operator in the Euclidean space with cylindrical coordinates $\left( \rho, \varphi, z \right)$. In terms of $\bm{\mathcal{E}}$ and $\bm \Phi$, the Einstein-Maxwell equations \eqref{eq:EM1} and \eqref{eq:EM2} can be written as the Ernst equations
    \begin{align}
        &\left( \operatorname{Re} \bm{\mathcal{E}} + |\Phi|^2 \right) \nabla^2 \bm{\mathcal{E}} = \left(\vec{\nabla} \bm{\mathcal{E}} + 2 \bm\Phi^* \vec{\nabla} \bm\Phi \right) \cdot \vec{\nabla} \bm{\mathcal{E}}, \label{eq:Ern1} \\
        &\left( \operatorname{Re} \bm{\mathcal{E}} + |\Phi|^2 \right) \nabla^2 \bm{\Phi} = \left(\vec{\nabla} \bm{\mathcal{E}} + 2 \bm\Phi^* \vec{\nabla} \bm\Phi \right) \cdot \vec{\nabla} \bm\Phi. \label{eq:Ern2}
    \end{align}

    The spacetime used in this work is generated by applying the Lie point symmetries of the Ernst equations to the seed Schwarzschild-Bertotti-Robinson metric, which is the non-rotating limit of the Kerr-Bertotti-Robinson black hole \cite{Podolsky:2025tle}. Its metric is given by 
    \begin{align} \label{eq:SBR}
        \rmd s^2 = \frac{1}{\Omega^2} \left[ - \mathcal{Q} \rmd t^2 + \frac{1}{\mathcal{Q}} \rmd r^2 + r^2 \left( \frac{1}{P} \rmd \theta^2 + P \sin^2 \theta \Delta_\varphi^2 \rmd \varphi^2 \right) \right],
    \end{align}
    with
    \begin{align}\label{eq:SBR_co}
        &P (\theta) = 1 + M^2 B^2 \cos^2 \theta, \nonumber \\
        &\mathcal{Q} (r) = \left( 1 + B^2 r^2 \right) \left( 1 - M^2 B^2 - \frac{2 M}{r} \right),  \\
        &\Omega^2 (r,\theta) = 1 + B^2 r^2 -B^2 r \left( r - 2 M - M^2 B^2 r \right) \cos^2 \theta. \nonumber
    \end{align}
    Here $M$ denotes the black hole mass and $B$ measures the strength of the Bertotti-Robinson magnetic field. The constant $\Delta_\varphi$ is a conicity factor. In particular, one may absorb it into the periodicity of the azimuthal angle $\varphi$. In this paper we keep the standard period $\varphi \in [0, 2\pi)$ by choosing 
    \begin{align}\label{eq:azi}
		\Delta_\varphi = \frac{1}{1 + M^2 B^2}. 
	\end{align}
    
    To apply the solution generating techniques, the seed metric must be cast into the Weyl-Lewis-Papapetrou form. Introducing
    \begin{align}\label{eq:Delta}
        \Delta_r (r) = r^2 \mathcal{Q}(r), \qquad \Delta_\theta (\theta) = P(\theta) \sin^2 \theta,
    \end{align}
    then Eq.~\eqref{eq:SBR} can be rewritten as
    \begin{align}\label{eq:SBR_WLP}
        \rmd s^2 = - f \left( \Delta_\varphi \rmd \varphi - \omega \rmd t \right)^2 + f^{-1} \left[ \rho^2 \rmd t^2 - e^{2 \gamma} \left( \frac{\rmd r^2}{\Delta_r} + \sin^2 \theta \frac{\rmd \theta^2}{\Delta_\theta} \right) \right],
    \end{align}
    where
    \begin{align}\label{eq:SBR_WLP_co}
        &f(r,\theta) = - \frac{r^2 \Delta_\theta}{\Omega^2}, \qquad \omega(r,\theta) = 0, \nonumber \\
		&\rho^2(r,\theta) = \frac{\Delta_r \Delta_\theta}{\Omega^4}, \qquad \gamma (r,\theta)= \frac{1}{2} \log \left[ \frac{r^4 \Delta_\theta}{\Omega^4}\right].
	\end{align}
    In this form, the magnetic potential of the seed metric is purely azimuthal. The functions $A_t$ and $h$ can be set to zero by a gauge choice. Therefore the seed Ernst potentials are
    \begin{align}\label{eq:seed_pot}
        \bm {\Phi}_0 = \frac{1 + M B^2 r \cos^2 \theta - \Omega}{B \Omega}, \qquad \bm{\mathcal{E}}_0 = \frac{2}{B} \bm{\Phi}_0.
    \end{align}
    
    The Harrison transformation can be used to add a Bonnor-Melvin magnetic field $b$ to the original metric \cite{Harrison:1968wue}, it acts on the Ernst potentials as 
    \begin{align}\label{eq:Harrison}
        &\bm {\mathcal{E}}_0 \rightarrow {\bm {\mathcal{E}}_{\mathrm H}} = \frac{\bm {\mathcal{E}}_0}{1 - 2 \alpha^* \bm {\Phi}_0 - |\alpha|^2 \bm {\mathcal{E}}_0}, \nonumber \\
        &\bm{\Phi}_0 \rightarrow {\bm \Phi}_{\mathrm H} = \frac{\bm {\Phi}_0 + \alpha \bm{\mathcal E}_0}{1 - 2 \alpha^* \bm {\Phi}_0 - |\alpha|^2 \bm {\mathcal{E}}_0}.
    \end{align}
    We only consider the pure magnetic case in order to keep the system as simple as possible. This can be achieved by taking only the real part of the complex Harrison transformation, namely $\alpha = b/2$.
    
    In addition, a swirling gravitational background can be obtained by combining the Harrison transformation with the Ehlers transformation \cite{Ehlers:1957zz}. For the purely magnetic seed metric considered here, the combined transformation can be written as 
    \begin{align}\label{eq:Ehlers}
        &\bm {\mathcal{E}}_0 \rightarrow \tilde {\bm {\mathcal{E}}} = \frac{\bm {\mathcal{E}}_0}{1 - 2 \alpha^* \bm {\Phi}_0 - |\alpha|^2 \bm {\mathcal{E}}_0 + i \jmath \bm {\mathcal{E}}_0}, \nonumber \\
        &\bm{\Phi}_0 \rightarrow \tilde{\bm \Phi} = \frac{\bm {\Phi}_0 + \alpha \bm{\mathcal E}_0}{1 - 2 \alpha^* \bm {\Phi}_0 - |\alpha|^2 \bm {\mathcal{E}}_0 + i \jmath \bm {\mathcal{E}}_0},
    \end{align}
    where $\jmath$ is the swirling parameter. The functions $\Delta_r$, $\Delta_\theta$$, \rho^2$ and $\gamma$ remain unchanged, while $f$ and $\omega$ are transformed into
    \begin{widetext}
		\begin{align}
			f(r,\theta) \rightarrow \tilde{f}(r,\theta) &= \frac{-4B^4 r^2 \Delta_\theta}{16 \jmath^2 \left( 1 + M B^2 r \cos^2\theta - \Omega\right) ^2 + \left[ b \left( b + 2B \right) \left( 1 + M B^2 r \cos^2\theta \right) - \left( b^2 + 2 b B + 2 B^2 \right) \Omega)\right] ^2}  , \label{eq:tf} \\  
			\omega(r,\theta) \rightarrow \tilde{\omega} (r,\theta) &= \frac{4 \jmath \left[ r - M \left( 2 + M B^2 r\right) \right] \cos \theta }{\Omega} + \omega_0. \label{eq:tomega}
		\end{align}
	\end{widetext}
    The integration constant $\omega_0$ corresponds to the choice of the frame, here we set $\omega_0 = 0$. The final metric is
    \begin{align}\label{eq:metric}
        \rmd s^2 = - \tilde{f} \left( \Delta_\varphi \rmd \varphi - \tilde\omega \rmd t \right)^2 + \tilde{f}^{-1} \left[ \rho^2 \rmd t^2 - e^{2 \gamma} \left( \frac{\rmd r^2}{\Delta_r} + \sin^2 \theta \frac{\rmd \theta^2}{\Delta_\theta} \right) \right].
    \end{align}
    The solution describes a Schwarzschild black hole embedded in a swirling Bertotti-Robinson-Bonnor-Melvin background. For $b=0$ or $b=-2B$, the additional Bonnor-Melvin contribution disappears, and the solution reduces to a Schwarzschild black hole in a swirling BR background. If one further sets $\jmath = 0$, the off-diagonal component $g_{t \varphi}$ vanishes and the spacetime reduces to the Schwarzschild-Bertotti-Robinson metric \cite{Podolsky:2025tle}. Another important case is $b=-B$, for which the external electromagnetic fields can be removed and the solution reduces to a static generalisation of the Schwarzschild black hole \cite{Astorino:2026okd}. The standard Schwarzschild geometry is recovered in the appropriate limit $B \to 0$. This limit has to be taken with care in the parametrization used above, since setting $B = 0$ directly in Eq.~\eqref{eq:tf} makes $\tilde{f}$ vanish, thus Eq.~\eqref{eq:metric} does not give a well-defined metric. 
	
	The event horizon is determined by $\Delta_r = 0$, namely
	\begin{align}\label{eq:EH}
		r_H = \frac{2 M}{1 - M^2 B^2}, 
	\end{align}
	Therefore, a regular event horizon requires $|B| M < 1$, and it approaches the Schwarzschild radius $r_S = 2 M$ in the weak-field regime $|B| M \ll 1$. 

    We now turn to the geodesic motion of particles. Since the particles we consider in this paper are neutral, they do not couple directly to the electromagnetic potential. The electromagnetic fields affect particle motion only through their backreaction on the spacetime geometry. From now on, we set the rest mass of the particle $m = 1$. The geodesic motion is generated by the Lagrangian
	\begin{align}\label{eq:Lag}
		\mathcal{L} = \frac{1}{2} g_{\mu \nu} \dot{x}^\mu \dot{x}^\nu, 
	\end{align}
	where an overdot denotes differentiation with respect to the proper time $\tau$. For timelike geodesics, the four-velocity $v^\mu = \dot{x}^\mu$ satisfies
    \begin{align}\label{eq:nor}
        g_{\mu \nu} v^\mu v^\nu = -1.
    \end{align}
    The corresponding canonical momentum is
	\begin{align}\label{eq:mom}
		p_\mu \equiv \dfrac{\partial \mathcal{L}}{\partial \dot{x}^\mu} =  g_{\mu \nu} \dot{x}^\nu, 
	\end{align}
	and the Hamiltonian is obtained by the Legendre transformation
	\begin{align}\label{eq:hal}
		\mathcal{H} \equiv p_\mu \dot{x}^\mu - \mathcal{L} = \frac{1}{2} g^{\mu \nu} p_\mu p_\nu.
	\end{align}
	Because the spacetime is stationary and axisymmetric, it admits two Killing vector fields $\partial_t$ and $\partial_\varphi$. They give two conserved quantities, the energy $E$ and $z$-component azimuthal angular momentum $L$ 
	\begin{align}
		E = - p_t =  -  g_{tt} \dot{t} - g_{t\varphi} \dot{\varphi} , \qquad L = p_\varphi = g_{t\varphi} \dot{t} + g_{t\varphi} \dot{\varphi}, \label{eq:con}
	\end{align}
	Solving Eq.~\eqref{eq:con} yields two first-order ordinary differential equations for $t$ and $\varphi$
	\begin{align}
		\dot{t} = \frac{{g}_{\varphi\varphi} E+{g}_{t \varphi} L}{{g}_{t \varphi}^{2}-{g}_{t t} {g}_{\varphi\varphi}}, \qquad \dot{\varphi} = -\frac{{g}_{t \varphi} E+{g}_{t t} L}{{g}_{t \varphi}^{2}-{g}_{t t} {g}_{\varphi\varphi}}, \label{eq:ODE1} 
	\end{align}
    The remaining equations of motion for $r$ and $\theta$ are 
    \begin{align}
		\ddot{r} =& \frac{1}{2} g^{rr} \left(\partial_r g_{t t} \dot{t}^{2} - \partial_r g_{r r} \dot{r}^{2} + \partial_r g_{\theta \theta} \dot{\theta}^{2} + \partial_r g_{\varphi\varphi} \dot{\varphi}^{2} \right.  \nonumber \\
		& \left. + 2 \partial_r g_{t \varphi} \dot{t} \dot{\varphi} - 2 \partial_\theta g_{rr} \dot{r} \dot{\theta}\right), \label{eq:ODE2} \\	
		\ddot{\theta} =& \frac{1}{2} g^{\theta \theta}\left( \partial_\theta g_{t t} \dot{t}^{2} + \partial_\theta g_{r r} \dot{r}^{2} - \partial_\theta g_{\theta \theta} \dot{\theta}^{2} + \partial_\theta g_{\varphi\varphi} \dot{\varphi}^{2} \right. \nonumber \\
		&+ \left.  2 \partial_\theta g_{t \varphi} \dot{t} \dot{\varphi} - 2  \partial_r g_{\theta \theta} \dot{r} \dot{\theta}\right). \label{eq:ODE3}
	\end{align}
	In addition, the normalization condition of the four-velocity also yields the constraint equation 
	\begin{align}
		\dot{r}^{2} + \frac{g_{\theta \theta}}{g_{r r}} \dot{\theta}^2 + V_{\mathrm{eff}} (r, \theta) = 0 , \label{eq:cons_eq}
	\end{align}
	where the two-dimensional effective potential is
	\begin{align}\label{eq:eff_pot}
		V_{\mathrm{eff}} (r, \theta) = \frac{1}{g_{r r}} \left[ 1 + \frac{{g}_{\varphi\varphi} E^2 + {g}_{t t} L^2 + 2 {g}_{t \varphi} E L}{{g}_{t t} {g}_{\varphi\varphi} - {g}_{t \varphi}^{2}} \right].
	\end{align}
	Eq.~\eqref{eq:cons_eq} determines the allowed region for particle motion and is useful for locating the boundary of the accessible domain on the Poincar\'{e} section, whose explicit discussion is given in Appendix~\ref{app:1}. Since no additional constant of motion similar to the Carter constant is known for this spacetime, the $r$ and $\theta$ motions remain coupled in general. This suggests that the geodesic Eqs.~\eqref{eq:ODE1}--\eqref{eq:cons_eq} are not expected to be separable and the possible non-integrability of the particle motion will be examined numerically below. In the next section, we investigate how the phase space structure of bound orbits changes when one parameter is varied while the others are fixed.

	\section{Chaotic behavior of particles}\label{sec:3}
	
	In this paper, we solve the geodesic equations numerically by using the corrected fifth-order Runge-Kutta method \cite{Wu:2007crk,Ma:2008cke,Ma:2008cli}. In this integrator, the velocity components $( \dot{r},\dot{\theta} ) $ are corrected at each integration step, and the accumulated numerical deviation is projected back along a least-squares shortest path. This procedure efficiently suppresses long-term error growth and helps to avoid pseudo chaotic behavior generated purely by numerical drift, see Appendix \ref{app:2}. 
	
	We require that the particle motion is not restricted to the equatorial plane. In principle, the three background parameters $B$, $b$ and $\jmath$ can vary independently, while the initial conditions of particles must satisfy the constraint equation \eqref{eq:cons_eq}. Throughout this paper, we fix the black hole mass $M = 1$. In the present numerical analysis, stable bound orbits do not exist once the external magnetic field parameters become sufficiently large. Therefore, we mainly focus on the weak-field regime $|B| M \ll 1$, which is also related to realistic astrophysical observations \cite{EventHorizonTelescope:2019dse,EventHorizonTelescope:2021srq,EventHorizonTelescope:2022wkp}. Indeed, even the strong magnetic fields inferred near supermassive black holes in galactic centers correspond to field strengths far below the characteristic scale $M^{-1}$ \cite{Wang:2025vsx}. Since the metric remains invariant under the transformation $(b,B) \rightarrow (-b,-B)$, we can choose $B > 0$ without loss of generality. With this convention, $b > 0$ means the two electromagnetic fields are aligned, whereas $b<0$ means that they are oppositely directed. 
	
	\subsection{Poincar\'{e} section}\label{sec:3.1}
	
    \begin{figure*}[tbph]
		\centering
		\includegraphics[width=\textwidth]{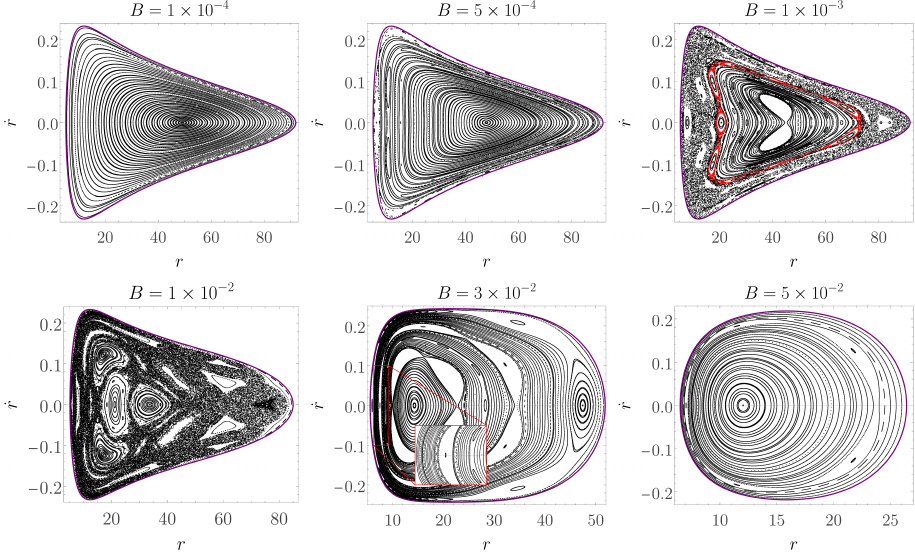}
		\caption{The Poincar\'{e} sections on the equatorial plane $(\theta = \pi/2)$ for different values of the Bertotti-Robinson parameter $B$ in the non-swirling case $\jmath = 0$. Here we fix the parameters $E = 0.99, L = 4.0, b = 0$ and initial conditions $\dot{r}(0) = 0, \theta(0) = \pi/2 $. The purple curve denotes the boundary of the accessible region.} 
		\label{fig:1} 
	\end{figure*}
    
	\begin{figure*}[tbph]
		\centering
		\includegraphics[width=\textwidth]{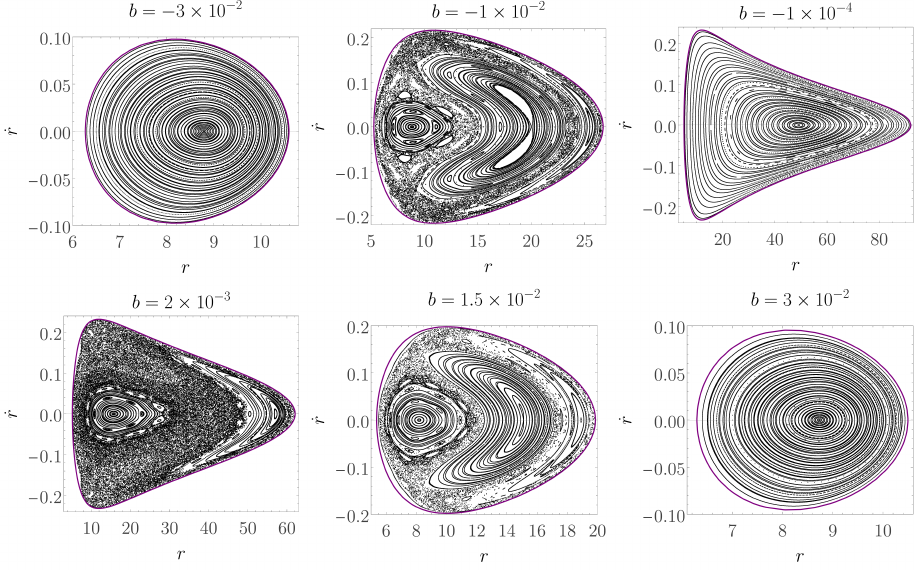}
		\caption{The Poincar\'{e} sections on the equatorial plane $(\theta = \pi/2)$ for different values of the Bonnor-Melvin parameter $b$ with $B = 1 \times 10^{-4}$ and $\jmath = 0$. The fixed parameters are $E = 0.99$, $L = 4.0$, with initial conditions $\dot{r}(0) = 0$ and $\theta(0) = \pi/2 $. The figure illustrates the non-monotonic dependence of the phase space structure on $b$. } 
		\label{fig:2} 
	\end{figure*}
	
	\begin{figure*}[tbph]
		\centering
		\includegraphics[width=\textwidth]{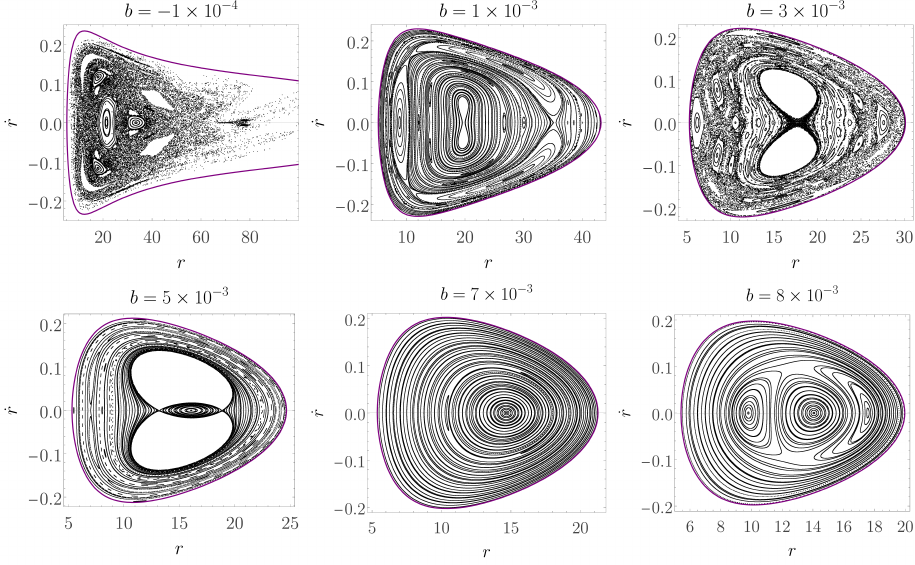}
		\caption{The Poincar\'{e} sections on the equatorial plane for different values of $b$ but with fixed $B = 1 \times 10^{-2}$. The remaining parameters and initial conditions are the same as in Fig.~\ref{fig:2}. The allowed region is more strongly constrained in this case, and stable bound orbits are almost absent for large negative values of $b$. } 
		\label{fig:3} 
	\end{figure*}
	
	\begin{figure*}[tbph]
		\centering
		\includegraphics[width=\textwidth]{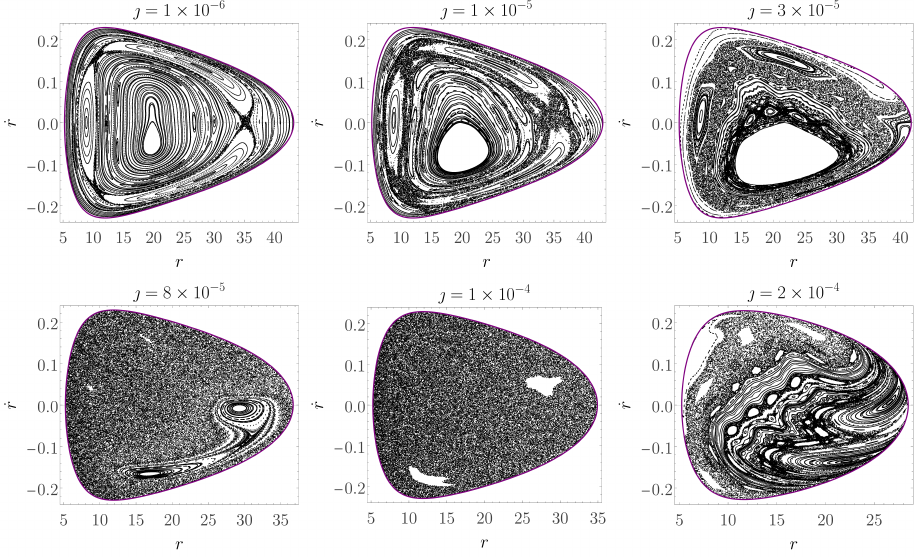}
		\caption{The change of Poincar\'{e} sections on the equatorial plane $(\theta = \pi/2)$ for different values of swirling parameter $\jmath$, while the other parameters are $E = 0.99$, $L = 4.0$, $B = 1 \times 10^{-2}$, $b = 1 \times 10^{-3}$, $\dot{r}(0) = 0$ and $\theta(0) = \pi/2 $. } 
		\label{fig:4} 
	\end{figure*}
	
	The Poincar\'{e} section is a standard tool for studying nonlinear Hamiltonian systems. It is constructed by selecting a codimension-one hypersurface in phase space and recording the intersection points each time an orbit crosses this hypersurface in a specified direction. In this way, the presence of chaotic motion can be identified directly from the structure of the Poincar\'{e} section: regular motion is represented by smooth invariant curves or resonant island chains, while the chaotic motion fills a finite area of the accessible domain densely. In this paper, we construct the Poincar\'{e} section on the equatorial plane $\left( \theta = \pi/2 \right) $, and record only the crossings satisfying $\dot{\theta} > 0$. 
	
	To separate the effect of the swirling background from the geometric deformation induced by the external electromagnetic fields, we first consider the non-swirling limit $\jmath = 0$. The other parameters are set to $E = 0.99$, $L = 4.0$, $b =  0$ to examine the effect of parameter $B$. In this case, the spacetime reduces to the Schwarzschild-Bertotti-Robinson black hole \cite{Podolsky:2025tle}. The initial conditions are chosen as $\dot{r}(0) = 0$ and $\theta(0) = \pi / 2$, then $\dot{\theta}(0)$ is calculated by Eq.~\eqref{eq:cons_eq}. 
    
    The resulting Poincar\'{e} sections are shown in Fig.~\ref{fig:1}. They indicate that the influence of parameter $B$ is non-monotonic. For a very weak field, such as $B = 1 \times 10^{-4}$, the Poincar\'{e} section is almost entirely filled by smooth curves surrounding a stable fixed point. This means that the motion remains close to integrable. The global structure is still largely regular when $B = 5 \times 10^{-4}$, although some invariant tori near the boundaries of the accessible region (drawn by the purple curve) begin to deform and the resonant structures become more visible. This suggests that the Bertotti-Robinson field has already started to weaken the original integrable structure. 
	
	When $B$ increases to $1 \times 10^{-3}$, a significant fraction of the Kolmogorov-Arnold-Moser (KAM) tori are destroyed and chaotic layers start to emerge, although there still exist some regular and resonance islands. In particular, the highlighted red orbit provides an example of weak chaos, since it only fills very thin layers associated with the separatrix of a resonant island chain. In nearly integrable Hamiltonian systems, such separatrix layers indicate that chaos will appear. For $B = 1 \times 10^{-2}$, the Poincar\'{e} section also exhibits a complicated structure composed of broken invariant curves, resonance islands and chaotic seas. However, the growth of chaos does not continue indefinitely. When $B = 3 \times 10^{-2}$, the chaotic behavior is obviously reduced and the phase space becomes more regular. The Poincar\'{e} section is again dominated by a series of invariant curves at $B = 5 \times 10^{-2}$, leaving only a few resonant islands. This indicates a regularization of particle motion. Therefore, the Bertotti-Robinson parameter $B$ does not simply enhance chaos. Instead, chaos appears in an intermediate range. Meanwhile, the accessible domain on the Poincar\'{e} section is obviously reduced as $B$ increases, showing that parameter $B$ affects not only the non-integrability of the system, but also the global geometry of the allowed phase space. 
	
	We next consider the influence of the Bonnor-Melvin parameter $b$. Its effect is more complicated and depends sensitively on the value of $B$. For this reason, we examine two representative values of parameter $B$, as shown in Figs.~\ref{fig:2} and \ref{fig:3}. First, we choose $B = 1 \times 10^{-4}$, because the spacetime is still close to integrable at $b = 0$ and stable bound orbits exist over a wide interval of both negative and positive values. Fig.~\ref{fig:2} shows the corresponding Poincar\'{e} sections for several representative $b$, while the remaining parameters are kept the same as before. Starting from a regular configuration at sufficiently large negative $b$, the phase space becomes more chaotic as $b$ approaches an intermediate value. Near $b = 0$, the motion becomes nearly regular again. For intermediate positive values, such as $b = 2 \times 10^{-3} $, a strong chaotic sea appears. When $b$ further increases, the motion becomes regular once more. Therefore, changing $b$ can drive the phase space through several transitions between regular and chaotic regimes. 
	
	To show that the above behavior is not restricted to the case where the initial state is nearly integrable, we also present a complementary example with $B = 1 \times 10^{-2}$ in Fig.~\ref{fig:3}. In this case, the motion is already chaotic at $b = 0$, as shown in the bottom left panel of Fig.~\ref{fig:1}. In addition, when the two electromagnetic fields are in opposite directions, stable bound orbits survive only in a very narrow range. Most trajectories escape after producing only a few points on the Poincar\'{e} section. As mentioned above, the accessible range of $b$ is strongly constrained by parameter $B$. When $b$ increases, the phase space first enters a regular regime at $b = 1 \times 10^{-3}$, then enters a strongly chaotic regime at intermediate values, finally becomes regular again for larger $b$. In other words, the effect of the Bonnor-Melvin parameter cannot be simply characterized by a monotonic trend, it depends on both the strength of the Bertotti-Robinson field and their relative orientation. In this sense, the Bonnor-Melvin parameter introduces richer characteristics of phase space and modifies the parameter regions in which chaos can occur. 
	
	Finally, we focus on the effect of the swirling parameter $\jmath$. In Fig.~\ref{fig:4}, we display the Poincar\'{e} sections for different values of $\jmath$ by choosing $B = 1 \times 10^{-2}$ and $b =  1 \times 10^{-3}$. We can find the phase space is still dominated by invariant curves and regular island chains for small $\jmath$. When $\jmath = 8 \times 10^{-5}$ or $1 \times 10^{-4}$, most of the KAM tori are destroyed and the Poincar\'{e} sections are almost occupied by a chaotic sea, with only a few isolated regular islands remaining. For $\jmath = 2 \times 10^{-4}$, the phase space topology becomes highly irregular and asymmetric. We also find that no stable bound orbits are obtained when $\jmath$ continues to increase. These results indicate that the main effect of the swirling parameter $\jmath$ is to drive the system toward a strongly non-integrable regime. Although the critical value at which the torus disappears depends on the background parameters and the chosen initial conditions, the qualitative effect of $\jmath$ is generally consistent. Since chaotic motion already exists in the non-swirling limit, $\jmath$ should not be regarded as the origin of chaos. Its main role is to modify the global phase space structure, accelerate the destruction of invariant tori and increase the probability of chaotic motion.

	\subsection{Lyapunov Exponents}\label{sec:3.2}

    \begin{figure*}[tbph]
		\centering
		\includegraphics[width=\textwidth]{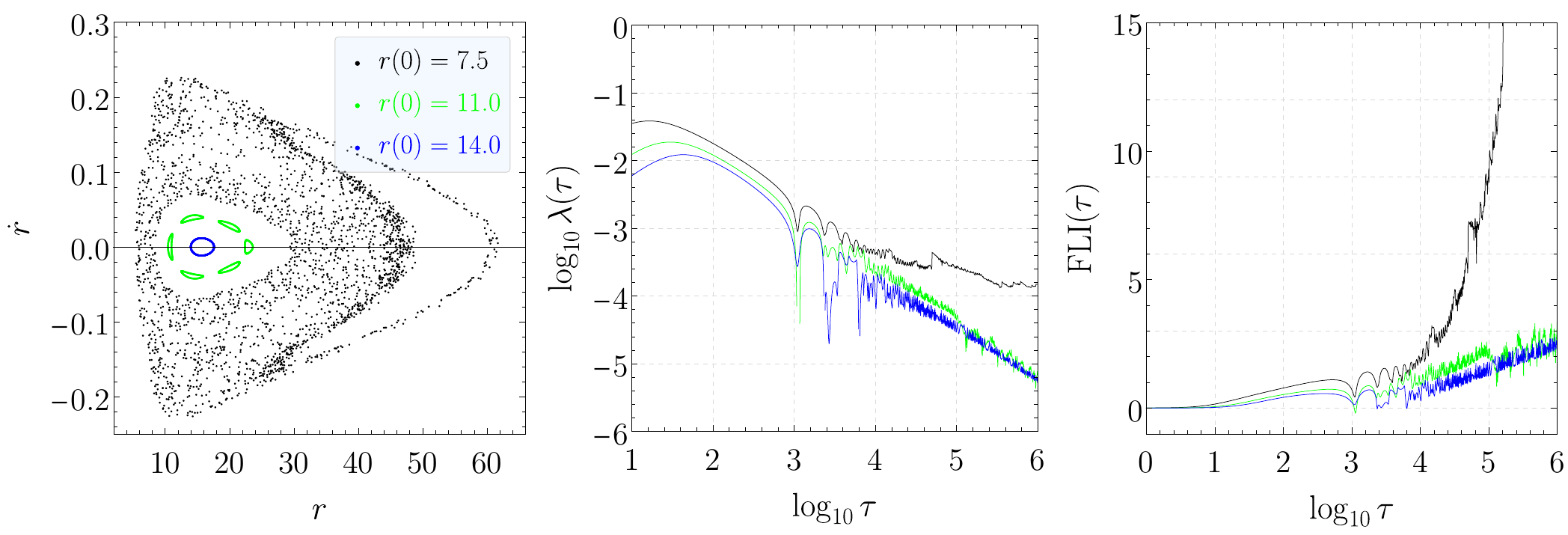}
		\caption{Left: Poincar\'{e} sections for three representative orbits with different positions $r(0) = 7.5$ (black), $11.0$ (green) and $14.0$ (blue), the background parameters are set to $B = 1 \times 10^{-4}$, $b = 2 \times 10^{-3}$ and $\jmath = 0$. Middle: evolution of the maximum LE as a function of the proper time $\tau$ in a log-log scale. Right: the corresponding FLI as a function of $\log_{10} \tau$. } 
		\label{fig:5} 
	\end{figure*}

	\begin{figure*}[tbph]
		\centering
		\includegraphics[width=\textwidth]{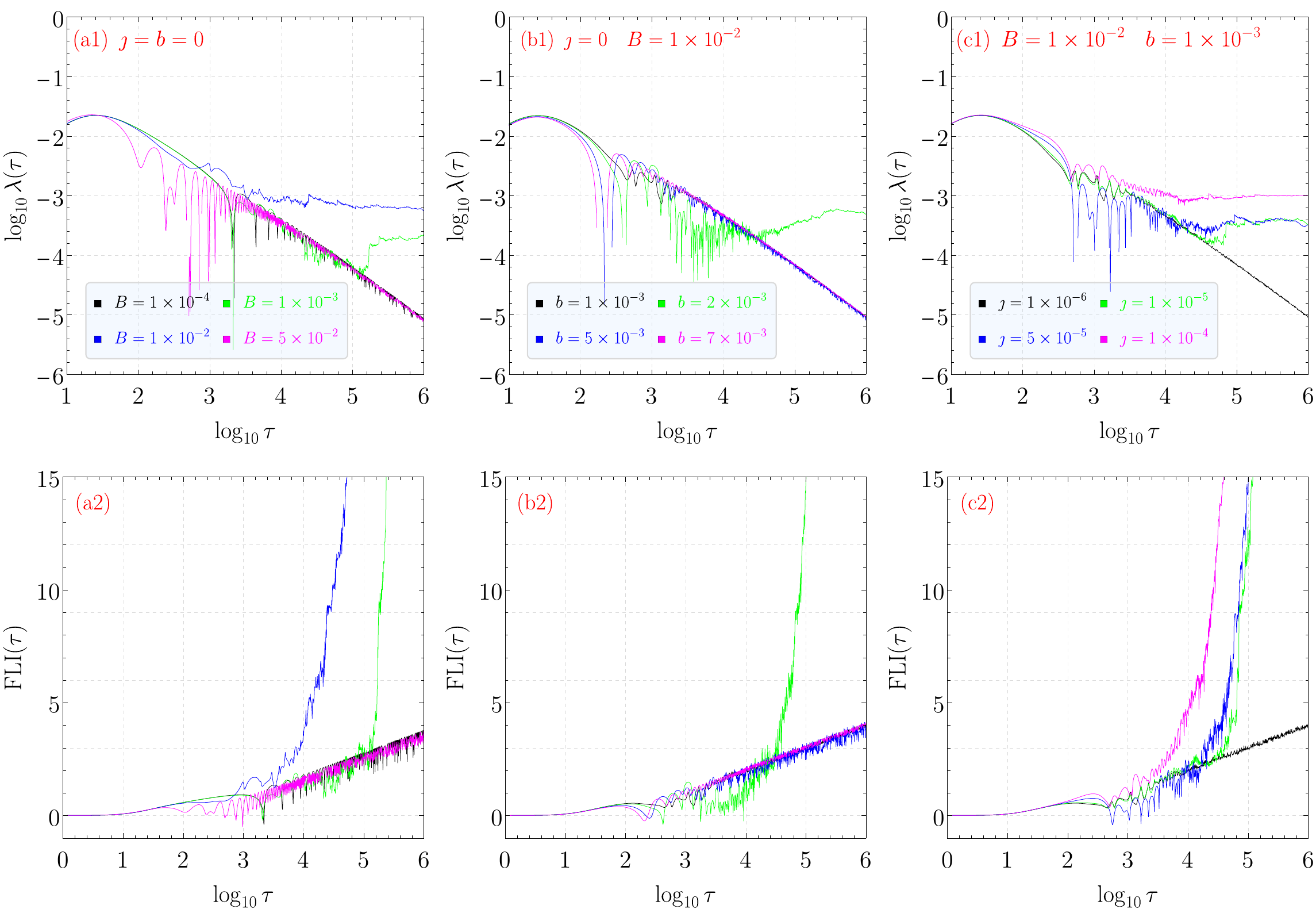}
		\caption{Upper panels: evolution of the maximum LEs for different trajectories with the same initial position $r(0) = 10.0$ in log-log scale. Lower panels: the corresponding FLIs. In the first column, $\jmath = b = 0$ and $B$ is varied. In the second column, $\jmath = 0$, $B = 1\times 10^{-2}$ and $b$ is varied. In the third column, $B = 1 \times 10^{-2}$, $b = 1\times 10^{-3}$ and $\jmath$ is varied. } 
		\label{fig:6} 
	\end{figure*}
    
	In this paper, we mainly use two quantitative indicators to measure the divergence of nearby trajectories. The first one is the maximum Lyapunov exponent (LE), following the procedure proposed by Refs.~\cite{Wu:2003pe,Wu:2006rx}, it is defined as
	\begin{align}
		\lambda &= \lim_{\tau \to \infty} \frac{1}{\tau} \ln \frac{d (\tau) }{ d (0)}, 
	\end{align}
	where $d (\tau)$ represents the proper distance between two nearby trajectories. In curved spacetime, this distance can be evaluated by projecting the deviation vector onto the local rest frame
	\begin{align}
		d (\tau)  = \sqrt{ \left| h_{\alpha\beta} \Delta x^\alpha \Delta x^\beta \right|},
	\end{align}
	where the projection tensor is
	\begin{align}
		h_{\alpha\beta} = g_{\alpha\beta} + v_\alpha v_\beta,
	\end{align}
	$v^\alpha$ is the four-velocity along the reference trajectory. The deviation vector is defined as 
	\begin{align}
		\Delta x^\alpha (\tau) = \tilde{x}^\alpha (\tau) - x^\alpha (\tau),
	\end{align}
	where $x^\alpha (\tau)$ and $\tilde{x}^\alpha (\tau)$ are evolved from two neighboring trajectories with slightly different initial positions. The norm of this deviation vector gives the instantaneous proper separation between the two trajectories. In the actual numerical calculation, we take a large but finite proper time as $\tau = 10^6 $ and adopt the following form
	\begin{align}
		\lambda (\tau) &=  \frac{1}{\tau} \sum_{i}^{N(\tau)} \ln \frac{d_i}{ d_0 }, 
	\end{align}
	where $d_0$ is the initial separation and $d_i$ is the separation immediately before the $i$-th renormalization. Such a renormalization is necessary since it can avoid numerical saturation caused by true exponential divergence. We set $ d_0 = 1 \times 10^{-9}$ and plot the maximum LE in a log-log scale. For regular motion, $\lambda(\tau)$ decreases as $\tau^{-1}$ and tends to zero, whereas for chaotic motion it approaches a nonzero value.
	
	Since the maximum LE may converge slowly, especially for weakly chaotic trajectories, a long integration time is often required. Therefore, we introduce the Fast Lyapunov indicator (FLI), which is more efficient to identify regular and chaotic motion \cite{Wu:2003pe,Wu:2006rx,Froeschle:1997fli,Froeschle:2005dds,Lukes-Gerakopoulos:2013qva}. The FLI is defined as
	\begin{align}
		\mathrm{FLI} (\tau) &= - k \left[ 1 + \log_{10} d(0) \right] + \log_{10} \frac{d (\tau) }{ d (0)}, 
	\end{align}
	where $k$ is the number of renormalization. Once $d(\tau) = 1$, we reset the value of $d(\tau)$ to $d(0)$ and increase $k$ by $1$. In the present calculation, we use the same initial separation $d_0 = 1 \times 10^{-9}$. For a chaotic orbit, the FLI grows much faster than it does for a regular orbit. Therefore, it usually separates regular and chaotic trajectories more rapidly. 
	
	Fig.~\ref{fig:5} shows the Poincar\'{e} section together with the corresponding maximum LE and FLI for three orbits with the same global parameters but different initial positions, $r(0) = 7.5$ (black), $11.0$ (green) and $14.0$ (blue). The background parameters are set to $B = 1 \times 10^{-4}$, $b = 2 \times 10^{-3}$ and $\jmath = 0$. The left panel displays their different behavior on the Poincar\'{e} section. The black orbit is irregularly distributed over a large area of the accessible phase space, indicating strong chaos. The green orbit is composed of six secondary KAM tori belonging to the same island chain, which is associated with the $1/6$ resonance. The blue orbit is regular and remains confined to a single invariant torus. These conclusions are confirmed by the two quantitative indicators shown in the middle and right panels. In the middle panel, the black curve tends to a nonzero value after $\tau \sim 10^4$, whereas the other two curves decrease approximately linearly with $\log_{10} \tau$ and approach zero. This is consistent with the conclusion obtained from the Poincar\'{e} section in Fig.~\ref{fig:2}. In the right panel, it is obvious that the black curve grows much more rapidly than the other two curves, which is also a typical signature of chaotic motion.
	
	In addition, we select some trajectories with the same initial position $r(0) = 10.0$ under different background parameters and show their maximum LEs and FLIs in Fig.~\ref{fig:6}. In each column, the upper and lower panels correspond to the same parameter choices, so that the behavior of the two indicators can be compared directly. In the first column, $B$ is varied with fixed $\jmath = b = 0$. In the second column, $b$ is varied with fixed $\jmath = 0$ and $B = 1 \times 10^{-2}$. In the third column, $\jmath$ is varied with fixed $B = 1 \times 10^{-2}$ and $b = 1 \times 10^{-3}$. These results confirm the similar conclusions inferred from the Poincar\'{e} sections in Sec.~\ref{sec:3.1}. Therefore, the background electromagnetic parameters can drive a series of transitions in the system between regular and chaotic motion.

	\subsection{Recurrence analysis}\label{sec:3.3}
	
	\begin{figure*}[tbph]
		\centering
		\includegraphics[width=\textwidth]{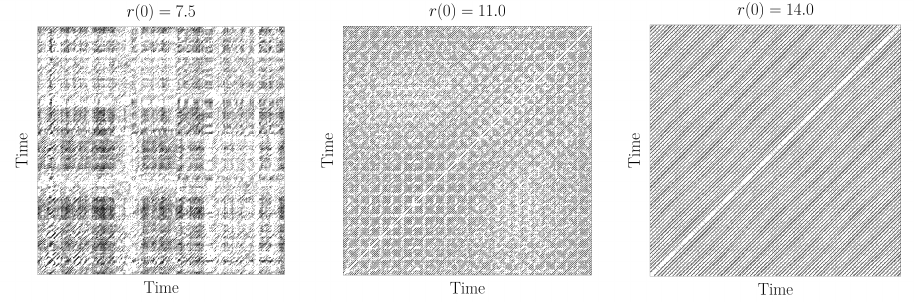}
		\caption{Recurrence plots corresponding to the three geodesic orbits shown in Fig.~\ref{fig:5}. The recurrence threshold is chosen separately for each orbit to ensure a fixed $\mathrm{RR} = 0.1$. Left: strongly chaotic orbit at $r(0) = 7.5$, using $\varepsilon = 1.7984$. Middle: resonant quasi-periodic orbit at $r(0) = 11.0$, using $\varepsilon = 1.0693$. Right: regular motion at $r(0) = 14.0$, using $\varepsilon = 0.5681$.} 
		\label{fig:7} 
	\end{figure*}
	
	Recurrence analysis is a powerful method for characterizing different dynamical behaviors of systems. It identifies the recurrences of a trajectory to neighborhoods of its previous states and it can clearly distinguish regular, chaotic and random evolutions \cite{Semerak:2012dx,Eckmann:1987rpd,Marwan:2007rpa,Kopacek:2010at,Kopacek:2010yr,Lukes-Gerakopoulos:2017jub}. Let $\vec{x}(i)$ denote a time series in the phase space sampled at equal proper time intervals, the recurrence matrix is
	\begin{align}
		\mathbf{R}_{ij} (\varepsilon) = \Theta \left( \varepsilon - \left\| \vec{x}(i) - \vec{x}(j)  \right\| \right) , \qquad i,j = 1,...,N.
	\end{align}
	Here $\varepsilon$ is the recurrence threshold, defined as the radius of the chosen neighborhood, $\Theta$ is the Heaviside step function, $N$ is the length of the trajectory. In practice, each coordinate time series is normalized separately to zero mean and standard deviation of one. 
	
	The recurrence plot provides a direct visualization of the recurrence matrix. Since $\mathbf{R}_{ij}$ takes only the values $0$ and $1$, a black dot is plotted when $\mathbf{R}_{ij} = 1$, while a white dot is left as $\mathbf{R}_{ij} = 0$. The main diagonal $\mathbf{R}_{ii} = 1$ is usually ignored. In the numerical implementation, we use the time delay method to reconstruct the phase space from the radial coordinate $r$, the reconstructed time series is 
    \begin{align}
        \vec{y}(t) = \left\{ \vec{x}(t) , \vec{x}(t + \Delta t ), ... , \vec{x}\left(t + (n - 1)\Delta t\right) \right\},
    \end{align}
    where $\Delta t$ is the time delay, which is usually chosen by the first minimum of the mutual information, while the embedding dimension $n$ can be determined by the false nearest neighbors method \cite{Kennel:1992na}. The length of the recurrence matrix is reduced to $N - (n - 1) \Delta t$. In this paper, we sample enough values of the radial coordinates $r$ along each orbit with constant time step $\Delta \tau$ and construct recurrence matrices with length $1000$.
    
    The simplest characteristic quantity of the recurrence matrix is the recurrence rate (RR), defined as the density of black points in the recurrence matrix 
	\begin{align}
		\mathrm{RR} (\varepsilon) = \frac{1}{N^2} \sum_{i = 1}^{N} \sum_{j = 1}^{N} \mathbf{R}_{ij} (\varepsilon). 
	\end{align}
    Fig.~\ref{fig:7} shows the recurrence plots for three orbits compared in Fig.~\ref{fig:5}. We select different recurrence thresholds to ensure the same recurrence rate $\mathrm{RR} = 0.1$. The left panel corresponds to a chaotic orbit, which has many vertical structures and shows highly irregular patterns. The right panel shows a clear long diagonal structures characteristic of regular motion. The middle panel represents a resonant quasi-periodic orbit, for which the diagonal structures are still visible but significantly weakened. 
    
    Although the recurrence plots already distinguish the dynamic behavior of different trajectories, one can further perform recurrence quantification analysis. For more details, see Ref. \cite{Marwan:2007rpa}. In what follows, we mainly consider the following quantifiers
    \begin{itemize}
        \item Determinism (DET), defined as the ratio of recurrence points forming diagonal lines longer than $l_{\mathrm{min}} = 2$.
        \item The longest diagonal line $L_{\mathrm{max}}$.
        \item Laminarity (LAM), defined as the ratio of recurrence points forming vertical structures longer than $v_{\mathrm{min}} = 2$. 
    \end{itemize}

    \begin{table}[]
	\centering
	\caption{Recurrence quantification measures for the three representative geodesic orbits shown in Fig.~\ref{fig:7}. The recurrence threshold $\varepsilon$, the time delay $\Delta t$ and the embedding dimension $n$ used in the time series reconstruction are also listed.}
	\label{tab:1}
	\setlength{\tabcolsep}{3mm}
	\renewcommand\arraystretch{1}
	\begin{tabular}{c|c|c|c|c|c|c}
		\hline 
		\hline
		$r(0)$ & $\varepsilon$ & $\Delta t$ & $n$ & $\mathrm {DET}$ & $L_{\mathrm{max}}$ & $\mathrm{LAM}$ \\
		\hline
		$7.5$ & $1.7984$ & $3$ & $5$ & $0.2285$ & $19$ & $0.3518$ \\
        \hline
		$11.0$ & $1.0693$ & $2$ & $5$ & $0.7604$ & $929$ & $0.1376$ \\
        \hline
		$14.0$ & $0.5681$ & $3$ & $3$ & $0.8756$ & $987$ & $0$ \\
		\hline
		\hline
	\end{tabular}
    \end{table}

    Tab.~\ref{tab:1} gives the quantitative values of the recurrence plots in Fig.~\ref{fig:7}. The strongly chaotic orbit with $r(0) = 7.5$ has small $\mathrm{DET}$ and $L_{\mathrm{max}}$, which means that its diagonal structures are short and strongly broken. In contrast, the resonant and regular quasi-periodic orbits have much larger values of $\mathrm{DET}$ and $L_{\mathrm{max}}$ because most points form parallel diagonal structures. However, the chaotic orbit has a larger value of $\mathrm{LAM}$ than the quasi-periodic cases, consistent with the vertical structures appearing in the recurrence plots. Therefore, recurrence analysis provides another effective way to distinguish different types of dynamical behaviors.

	\subsection{Bifurcation diagram}\label{sec:3.4}
	
	\begin{figure*}[tbph]
		\centering
		\includegraphics[width=\textwidth]{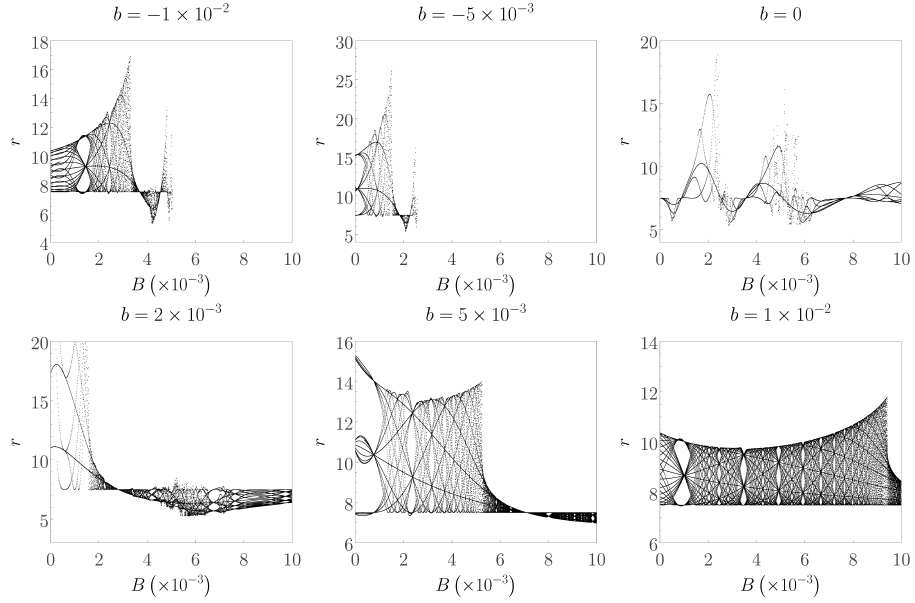}
		\caption{Bifurcation diagrams of the radial coordinate $r$ with respect to the Bertotti-Robinson parameter $B$ for different values of the Bonnor-Melvin parameter $b$ in the non-swirling case $\jmath = 0$. The constants of motion are $E = 0.99$ and $L = 4.0$, with initial conditions $r(0) = 7.5$, $\dot{r}(0) = 0$ and $\theta(0) = \pi/2 $. The blank intervals correspond to parameter regions where no long-lived bound orbit is obtained from the selected initial data. } 
		\label{fig:8} 
	\end{figure*}
	
	\begin{figure*}[tbph]
		\centering
		\includegraphics[width=\textwidth]{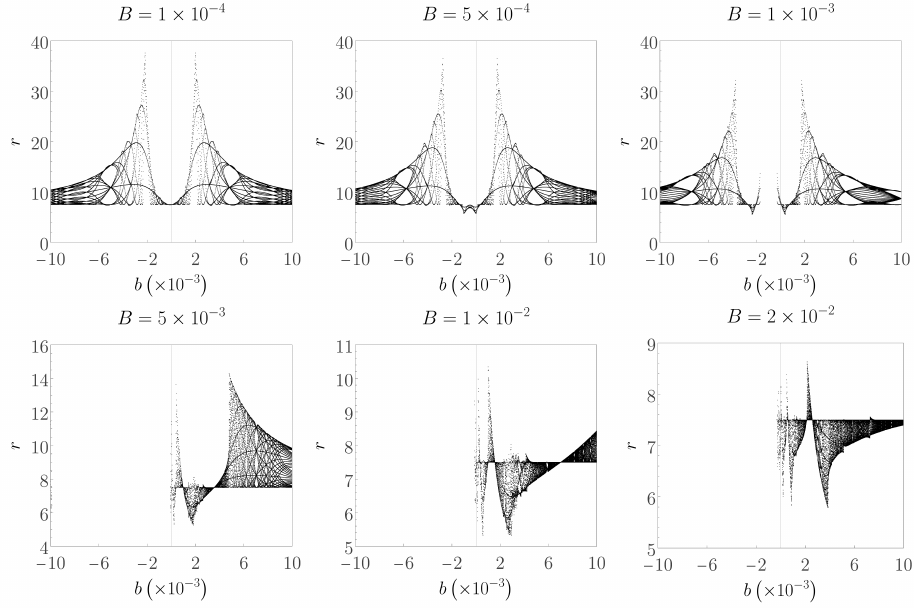}
		\caption{Bifurcation diagrams with respect to the Bonnor-Melvin parameter $b$ for different fixed values of $B$ in the non-swirling limit. The other parameters and initial conditions are the same as in Fig.~\ref{fig:8}. These diagrams show that the allowed range of bound orbits and the bifurcation structure are strongly controlled by the combined effect of two external fields.} 
		\label{fig:9} 
	\end{figure*}
	
	\begin{figure*}[tbph]
		\centering
		\includegraphics[width=\textwidth]{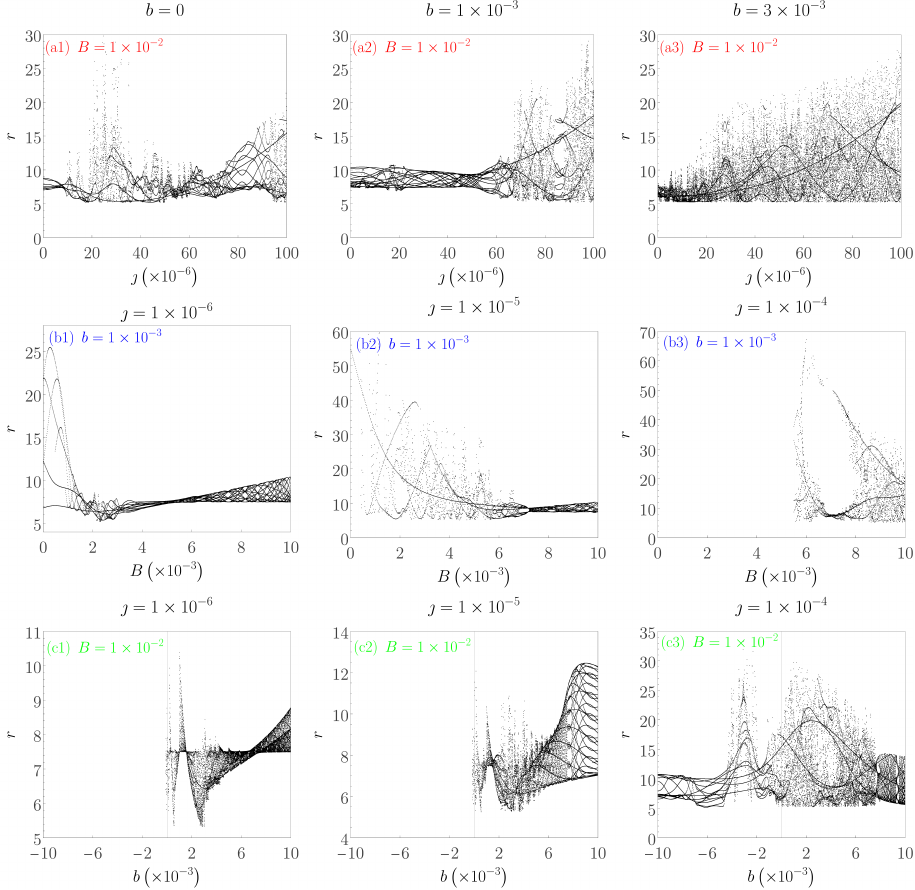}
		\caption{Bifurcation diagrams showing the influence of the swirling parameter $\jmath$. The upper row presents the bifurcation diagrams with respect to $\jmath$ for fixed $B = 1 \times 10^{-2}$ and different $b$. The middle row shows the bifurcation diagrams with respect to $B$ for fixed $b = 1 \times 10^{-3}$ and different $\jmath$. The lower row shows the bifurcation diagrams with respect to $b$ for fixed $B = 1 \times 10^{-2}$ and different $\jmath$. } 
		\label{fig:10} 
	\end{figure*}
	
	Finally, the bifurcation diagram provides a global view of how the dynamical behavior changes with a given parameter. It can be used to identify regions of parameter space in which chaos may occur. Regular regimes are characterized by smooth branches, which correspond to stable periodic or quasi-periodic motion. In contrast, densely and irregular distributions of points indicate chaotic dynamics. The appearance of periodic windows also suggests the regularization of the dynamics, while the disappearance of branches indicates that no stable bound orbits is obtained for the chosen parameters and initial conditions. 
	
	In Fig.~\ref{fig:8}, we present the bifurcation diagrams of the radial coordinate $r$ with respect to $B$ for different values of $b$ in the non-swirling case $\jmath = 0$. The energy and angular momentum are fixed as $E = 0.99$ and $L = 4.0$. The initial conditions are $r(0) = 7.5$, $\dot{r}(0) = 0$, $\theta(0) = \pi/2 $. These diagrams further support the non-monotonic influence of the external electromagnetic background on the system. In particular, the bifurcation diagram for $b = 0$ reproduces the  same trend already observed in Fig.~\ref{fig:1}. The system undergoes successive transitions among regular and chaotic motions as $B$ increases. Although the Bonnor-Melvin parameter $b$ also changes the dynamical behavior significantly, the same qualitative conclusion still remains. 
	
	Fig.~\ref{fig:9} presents the bifurcation diagrams of $r$ as a function of $b$ for several fixed values of $B$. It is obvious that changing $b$ also induces transitions between regular, chaotic and escaping regimes. For negative $b$, bound orbits rapidly disappear as $B$ increases, indicating that two oppositely directed electromagnetic fields tend to make the system unstable. This behavior agrees with the previous conclusion. The dense point distributions around intermediate values of $b$ correspond to the chaotic Poincar\'{e} sections in Fig.~\ref{fig:3}. For the selected initial data, most of the orbits will escape to infinity when $B$ continues to increase. Therefore, the combined effect of two electromagnetic fields is essentially nonlinear.
	
	In Fig.~\ref{fig:10}, we include the effect of swirling parameter $\jmath$. The upper row shows the bifurcation diagrams with respect to $\jmath$ for several $b$ at $B = 1 \times 10^{-2}$. When $b = 0$, the irregular regions are wider than the non-swirling case. For $b = 1 \times 10^{-3}$, the branches remain relatively narrow at very small $\jmath$, then become strongly scattered as $\jmath$ increases. This behavior is consistent with Fig.~\ref{fig:4}, where the Poincar\'{e} sections show a transition from regular islands to a large chaotic sea. In particular, the bifurcation diagram is irregular over a larger range at $b = 3 \times 10^{-3}$, implying that swirling can further enhance the nonintegrable behavior. Moreover, the middle and lower rows show how the $B$ and $b$ dependent bifurcation structures are modified when $\jmath$ is nonzero. Compared with Figs.~\ref{fig:8} and \ref{fig:9}, a finite swirling parameter shifts the locations of the regular and chaotic windows in general. For small $\jmath$, the global structure is similar to the non-swirling case. However, as $\jmath$ increases, the topological structure of phase space changes significantly, the chaotic regime expands and escape orbits become more frequent. Thus, the bifurcation diagram clearly confirms the previous results.

	\section{Conclusions}\label{sec:4}
	
	In this work, we have numerically studied the off-equatorial motion of particles around a Schwarzschild black hole immersed in a swirling Bertotti-Robinson-Bonnor-Melvin electromagnetic background. The main purpose was to clarify how the two different electromagnetic field components and the swirling deformation affect the integrability of particle motion around a non-rotating black hole. To this end, we combined several effective tools, including Poincar\'{e} sections, the maximum Lyapunov exponent, the Fast Lyapunov indicator, recurrence plots and bifurcation diagrams. 

    Our results show that the influence of the electromagnetic fields is very complicated. This behavior is different from a simple picture in which an electromagnetic field only enhances or suppresses chaos. In the present spacetime, the Bertotti-Robinson and Bonnor-Melvin fields modify both the characteristics of the effective potential and the accessible domain of phase space. When the two fields are oppositely directed, the range of stable bound motion can be strongly reduced. Therefore, changing these parameters can drive the system through successive transitions among regular, resonant and chaotic motion. This indicates that the geometric backreaction of the electromagnetic background plays an important role in breaking the integrable structure of particle motion. From the perspective of particle dynamics, our results reveal the nonlinear interaction between the two electromagnetic fields. 

    We also find that chaotic motion can appear in the non-swirling limit. In particular, even the Schwarzschild-Bertotti-Robinson spacetime can exhibit chaotic behavior. This indicates that the swirling background is not the only source for chaos in this family of black holes, the electromagnetic deformation of the spacetime geometry is already sufficient to produce non-integrable dynamics. The role of the swirling parameter is to accelerate the destruction of invariant tori, produce clear signatures of chaotic behavior and increase the possibility of escaping orbits. 

    The different methods used in this work give consistent conclusions. The Poincar\'{e} sections directly show the transition from invariant tori to resonant islands and chaotic seas. The maximum Lyapunov exponent and the fast Lyapunov indicator provide quantitative measures for analyzing the divergence of particle trajectories. The recurrence plots and recurrence quantification further reveal the recurrent structures in phase space, allowing us to distinguish different behaviors of regular, resonant and chaotic orbits. The bifurcation diagrams give a global view of the phase space transitions for different background parameters. These independent methods support the fact that the external electromagnetic background can reconstruct the phase space of particle motion. In summary, particle dynamics can be used as a powerful probe of how astrophysical environment that may exist in strong gravitational fields affects the phase space structure. 
	
	For future work, the motion of charged particles should be studied, since the Lorentz force will introduce a direct coupling between the particle and the electromagnetic potential. This may lead to stronger chaotic behavior and a more complicated phase space structure. The dynamics of spinning particles in this spacetime is also worth investigating. The spin-curvature coupling may provide another nonlinear interaction. Finally, it is of great interest to explore whether the chaos induced by the external background can leave observable signatures in black hole shadows, accretion dynamics or extreme mass ratio inspiral system around black holes. 
	
	\section{Acknowledgments}\label{sec:5}
	
	We would like to thank Surojit Dalui, Dan-Dan Yuan and Jia-Geng Jiao for their helpful discussions. W. Li would also like to thank Zhi-Chao Wang for his valuable suggestions. This work was partially supported by the National Natural Science Foundation of China (NSFC) (Grant Nos.~12275166 and 12311540141). YQL was partially supported by NSFC, China (Grant No.~12405072) and China Postdoctoral Science Foundation (Grant No.~2024M761914).

	\appendix
	
	\section{The effective potential} \label{app:1}

    \begin{figure}[tbph]
		\centering
		\includegraphics[width=0.4\textwidth]{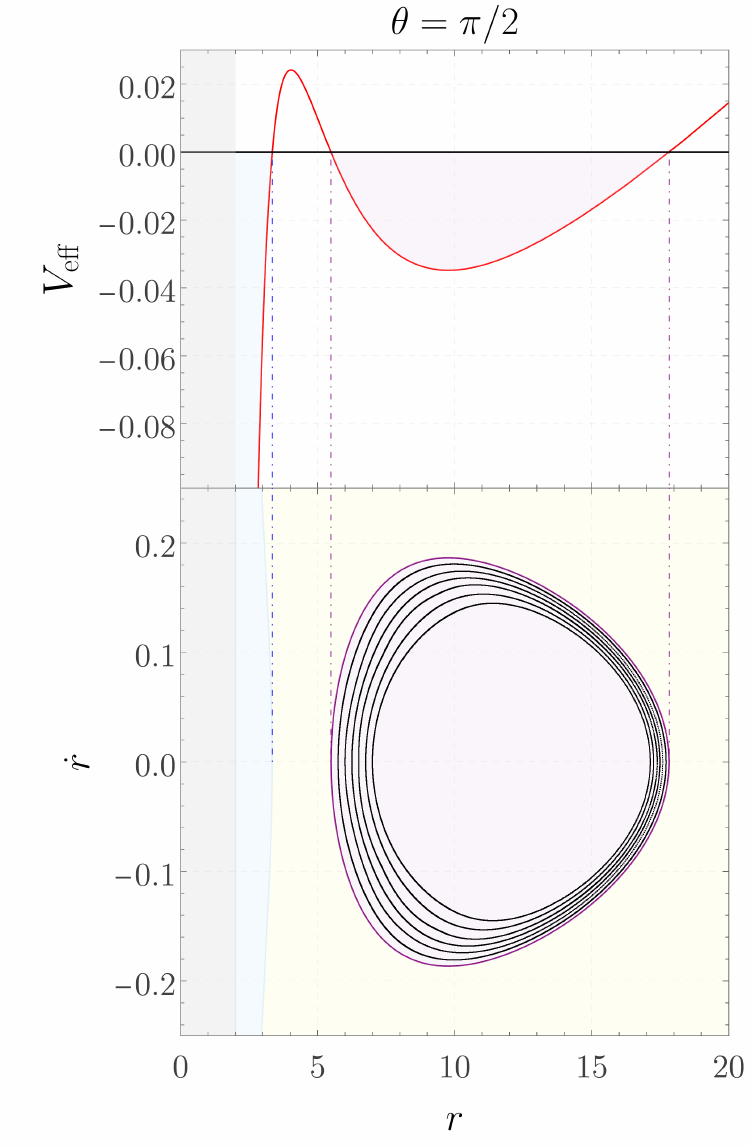}
		\caption{The effective potential and allowed domain on the equatorial Poincar\'{e} sections for $\jmath = 0$, $B = b = 1 \times 10^{-2}$, $E = 0.99$ and $L = 4.0$. The upper panel shows $V_{\mathrm{eff}} (r, \pi/2)$. The gray shaded region denotes the black hole interior. The roots of $V_{\mathrm{eff}}$ separate different intervals. The lower panel shows the projection of the allowed region onto the $(r, \dot{r})$ plane, together with six Poincar\'{e} sections at different radial positions. The blue shaded region is connected to the horizon and corresponds to plunging motion, while the purple curve gives the boundary of the outer allowed region. The yellow region is not physically allowed.} 
		\label{fig:app:1} 
	\end{figure}
    
    In this appendix, we clarify the role of the effective potential Eq.~\eqref{eq:eff_pot} in determining the allowed region of the particle motion. This discussion also provides a useful supplement to the Poincar\'{e} sections shown in the main text. Outside the event horizon, we have
    \begin{align}
        g_{rr} > 0, \qquad {g_{\theta\theta}} > 0,
    \end{align}
    Therefore, a necessary condition for particle motion is 
    \begin{align}
       r > r_H, \qquad V_{\mathrm{eff}} (r, \theta) \leq 0.
    \end{align}
    The contour $V_{\mathrm{eff}} (r, \theta) = 0$ then defines the curve of zero velocity in the $(r, \theta)$ configuration space. On this curve, the radial and polar velocity equal to zero at the same time. Hence, it separates the physically allowed region from the forbidden region in the meridional plane. The roots of $V_{\mathrm{eff}}$ determine the boundaries of plunging or bound orbits. 
    
    In the present work, the Poincar\'{e} section is chosen on the equatorial plane $(\theta = \pi/2)$, so it is useful to project the allowed domain onto the $(r, \dot{r})$ plane. For a given $r$, the maximum value of radial velocity is obtained by setting $\dot{\theta} = 0$ in Eq.~\eqref{eq:cons_eq}. This gives
    \begin{align}
        \dot{r}_\pm (r) = \pm \sqrt{- V_{\mathrm{eff}} (r, \pi/2)},
    \end{align}
    which means that the region between the two branches $\dot{r}_-(r)$ and $\dot{r}_+(r)$ is allowed. 

    As illustrated in Fig.~\ref{fig:app:1}, the upper panel displays the effective potential $V_{\mathrm{eff}} (r, \pi/2)$ for $\jmath = 0$, $B = b = 1 \times 10^{-2}$ with $E = 0.99$ and $L = 4.0$. The gray shaded region corresponds to the black hole interior $r < r_H$, and is excluded from the particle dynamics. In this case, two disconnected allowed regions appear outside the event horizon. The inner allowed region (blue) is connected to the event horizon and therefore corresponds to the plunging orbits. The outer region (purple) is separated from the horizon by a positive potential barrier and forms a potential well, in which bounded non-plunging orbits can exist. 

    The lower panel shows the corresponding projection of the allowed domain onto the $(r, \dot{r})$ plane. The blue shaded region gives plunging orbits, which indicates that initial conditions placed in this region are allowed but are expected to be captured by the black hole. The purple curve denotes the boundary of the outer allowed region on the Poincar\'{e} section. The interior of this curve is physically allowed and contains the bounded non-plunging orbits, which is also our main concern here. The yellow region is forbidden, because the points in this region would give $\dot{\theta}^2 < 0$. This will lead to a non-zero imaginary part of the polar velocity.

    \section{The constraint error of numerical integration} \label{app:2}

    \begin{figure}[htbp]
		\centering
		\includegraphics[width=0.4\textwidth]{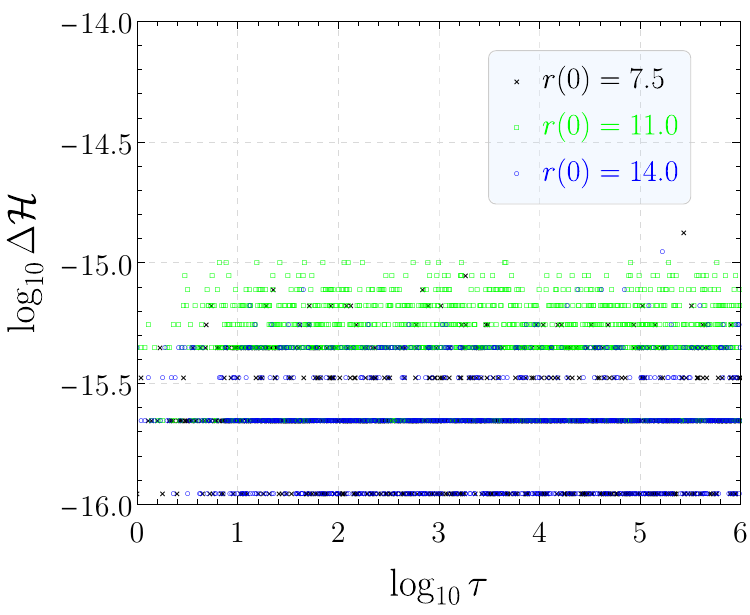}
		\caption{The relative error $\Delta \mathcal{H}$ for three orbits in Fig.~\ref{fig:5}. We find that the relative error remains below $10^{-14}$ and the numerical integration has long-term stability.} 
		\label{fig:app:2} 
	\end{figure}

    The Hamiltonian constraint should be preserved during the whole process of numerical evolution. To examine the reliability of the numerical integration, we monitor the relative error in Hamiltonian \eqref{eq:hal}, defined by 

    \begin{align}
        \Delta \mathcal{H}(\tau) = \left| 1 - \frac{\mathcal{H}(\tau)}{\mathcal{H}(0)} \right|,
    \end{align}
    In Fig.~\ref{fig:app:2}, we show $\log_{10} \Delta \mathcal{H}$ corresponding to the trajectories in Fig.~\ref{fig:5}. It can be seen that the relative error remains below $10^{-14}$ throughout the integration and does not exhibit any irregular growth. This demonstrates that the integrator used in this work maintains stable and high precision Hamiltonian constraint, so the dynamical structures identified in the main text should not be caused by the accumulation of numerical errors.


\end{document}